\documentclass[twocolumn,floats,aps,showpacs]{revtex4}
\usepackage{times,amsmath,graphics,psfrag,latexsym}

\begin{document}
\title{Pressure and linear heat capacity in the superconducting state of thoriated UBe$_{13}$}
\author{R.J. Zieve$^1$, R. Duke$^1$ and J.L. Smith$^2$}
\affiliation{${}^1$Physics Department, University of California at Davis\\
${}^2$Los Alamos National Laboratory, Division of Materials Science
and Technology}
\begin{abstract}
Even well below $T_c$, the heavy-fermion superconductor
(U,Th)Be$_{13}$ has a large linear term in
its specific heat.  We show that under uniaxial pressure, the
linear heat capacity increases in magnitude by more than a factor of two. The
change is reversible and suggests that the linear term is an intrinsic property of
the material.  In addition, we find no evidence of hysteresis or of latent heat in
the low-temperature and low-pressure portion of the phase diagram, showing that all
transitions in this region are second order.          
\end{abstract} 
\maketitle

For the past two decades, heavy-fermion superconductors have
revealed a variety of unusual behaviors that hint at unconventional
superconductivity \cite{Heffner96}.  Many low-temperature properties
such as specific heat and NMR relaxation rates have power law rather
than exponential temperature dependences. Two compounds, UPt$_3$ and
U$_{1-x}$Th$_x$Be$_{13}$ for $x$ between 0.02 and 0.04, each have two
transitions leading to distinct superconducting phases.  Further phase
transitions appear with pressure or applied magnetic field. Yet no
experiment has emerged that conclusively identifies the order parameters
of the phases or even their symmetry.

One potential clue to heavy-fermion order parameters is the
significant linear term in the thermal properties within the superconducting
state. In theory linear
heat capacity and thermal conductivity are normal-state phenomena, which
should disappear once the superconducting energy gap alters the excitation
spectrum. Yet specific heat $C(T)=\gamma_s T + C_{non-linear}(T)$
in the superconducting phase is a persistent feature in heavy-fermion
superconductors.  The coefficient $\gamma_s$ can reach over 50\%
of $\gamma_n$, the normal-state value of $C(T)/T$ just above the
transition. First seen in UPt$_3$ \cite{Fisher89}, a large $\gamma_s$
and an analogous linear term in thermal conductivity are also
found in CeCoIn$_5$ \cite{Movshovich}, URu$_2$Si$_2$ \cite{Fisher90,
Behnia92}, and UPd$_2$Al$_3$\cite{Caspary, Chiao97}, among others.

The linear term is sometimes viewed as stemming entirely or in part from
imperfect samples.  One simple explanation would be that some fraction
of the material remains normal.  Another heavily discussed alternative,
resonant impurity scattering, combines impurity effects with intrinsic
features of the order parameter \cite{Hirschfeld, Balatsky, Hass}.
For a $d$-wave superconductor, a very small impurity concentration could
create finite ungapped regions on the Fermi surface and a constant density
of states.  The strength, phase shift, and anisotropy of the impurity
scattering and the impurity density all factor into the normal-like
behavior.  Yet other scenarios treat the linear
term as an intrinsic property.  These include involvement of only part of the
Fermi surface in superconductivity, and the more exotic odd-frequency
pairing \cite{Coleman}.

In support of the importance of impurities, the magnitude of $\gamma_s$ varies
significantly from sample to sample, ranging from $0.12\gamma_n$ to
$0.62\gamma_n$ in UPt$_3$. Furthermore, $\gamma_s$ generally decreases as the
superconducting transition temperature of the sample increases.  Since a higher
$T_c$ often indicates a better-quality sample, the correlation does suggest
that $\gamma_s$ arises at least in part from sample problems \cite{Fisher89}.
Also, $\gamma_s$ in pure UBe$_{13}$ is small or zero, but doping with either
thorium \cite{Jin94} or boron \cite{Ott91, Beyermann} increases both $\gamma_s$
and the likelihood of sample inhomogeneities.

\begin{figure}
\begin{center}
\scalebox{0.7}{\includegraphics{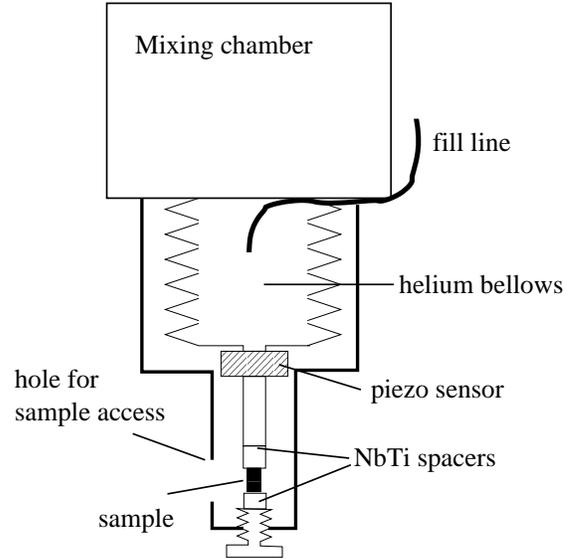}}
\caption{\small Helium bellows setup for measuring specific heat under
uniaxial pressure.}
\label{f:cell}
\end{center}
\end{figure}

\begin{figure}
\begin{center}
\psfrag{C/T (J/mol K^2)}{\scalebox{2}{$C/T$ (J/mol K$^2$)}}
\scalebox{0.45}{\includegraphics{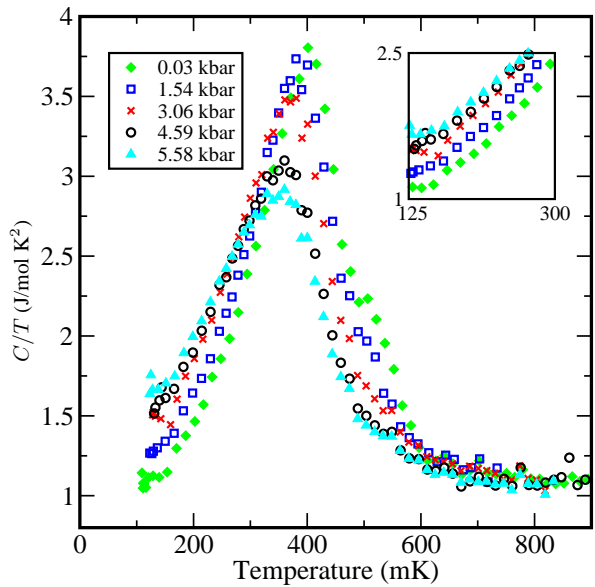}}
\caption{\small Specific heat of U$_{0.98}$Th$_{0.02}$Be$_{13}$ as a function
of temperature, for several applied uniaxial pressures.  Inset enlarges
part of the low-temperature region, showing that the curves for different
pressures are nearly parallel.}
\label{f:CvsT}
\end{center}
\end{figure}

Any theory of the linear terms in the superconducting
specific heat and thermal conductivity
must also address the absence of a linear contribution to the NMR
spin-lattice relaxation rate.  If the large $\gamma_s$ comes from either a
normal portion of the sample or a finite region of the Fermi surface with
no gap, that same source should lead to a linear Korringa relaxation,
as in the normal state.  However, NMR measurements on a variety of
heavy-fermion superconductors find only a cubic temperature dependence
down to temperatures well below where linear terms in heat capacity
and thermal conductivity become significant \cite{MacLaughlin, Matsuda,
Kohori}.  The cubic dependence would be expected from line nodes in
the gap.  Reconciling the NMR and thermodynamic data would certainly be
a step towards understanding heavy-fermion superconductivity.

Our present work shows a large and reversible change in $\gamma_s$ with
pressure in U$_{0.98}$Th$_{0.02}$Be$_{13}$.  Explaining the pressure
dependence, which occurs without a change in the impurity concentration,
will further restrict theoretical treatments.

We use a pressure cell activated by a helium bellows.  The setup is mounted at
the mixing chamber of a KelvinOx 100 dilution refrigerator, and we can change
pressure while keeping the sample temperature below 300 mK. Our cell,
illustrated schematically in Figure \ref{f:cell}, is modelled after the cell
described by Pfleiderer {\em et al.} \cite{Pfleiderer}. The expanding bellows
presses on a column including the sample and a piezoelectric crystal to measure
pressure changes. The small cross-sectional area of the sample amplifies the
pressure within the bellows; by the time the helium solidifies at 25 bar we
reach a uniaxial pressure of 7.8 kbar at the sample.   The uniaxial technique is
necessary for changing pressure at low temperature.  To avoid symmetry-breaking
effects from the uniaxial pressure, we use a polycrystalline sample in the
experiment.

We use a transient pulse method for heat capacity measurements.
Our heater is a 50:50 AuCr thin film, our thermometer a RuO$_2$
film. Pieces of NbTi on each side of the sample provide a thermal link,
with a time constant of order 8 seconds between the sample and the rest
of the bellows.  As a conventional superconductor well below $T_c$,
the NbTi itself contributes negligibly to the measured heat capacity.

One difficulty with the measurements is a significant time constant
between the helium bellows and the dilution refrigerator, of
order 3 minutes near 500 mK and increasing to 10 minutes at 200 mK.
Waiting for the bellows to equilibrate completely with the cryostat
at each temperature takes prohibitively long. Instead, we measure the
relaxation time of the bellows throughout our temperature range and verify
that it is independent of pressure.  We then account for a slowly changing
bellows temperature in our fits of the temperature decays after each heat pulse.

In Figure \ref{f:CvsT}, we show $C/T$ as a function of temperature
for five different pressures.  At the lowest pressure there are two
transitions, centered at 450 mK and 550 mK, with a small shoulder
between them. As pressure increases, the two transitions decrease in
temperature and merge into one, the shoulder disappearing.  The amplitude
of the peak decreases, while that of the low-temperature tail increases.
The normal-state heat capacity does not change with pressure. All this
agrees with the previous uniaxial pressure experiment \cite{Zieve94}.

\begin{figure}
\begin{center}
\psfrag{g0 (J/mol K^2)}{\scalebox{2}{$\gamma_s$ (J/mol K$^2$)}}
\psfrag{10^5 A}{\scalebox{1.8}{$10^5 A$}}
\scalebox{0.45}{\includegraphics{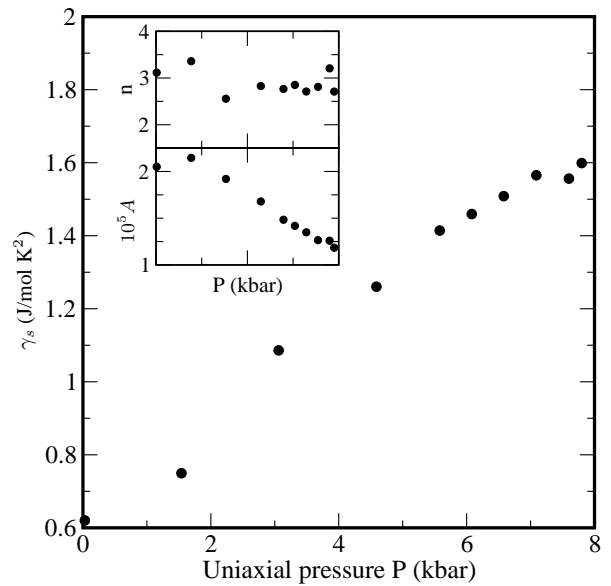}}
\caption{\small Linear coefficient of specific heat in superconducting
phase, $\gamma_s$, as a function of pressure. 
Inset: Best fit parameters for low temperature tail: the exponent
$n$ from a 3-parameter fit $C(T)=\gamma_s T+AT^n$ (top) and the prefactor
$A$ from a 2-parameter fit $C(T)=\gamma_s T+AT^3$ (bottom).}
\label{f:gamma}
\end{center}
\end{figure}

The lowest-pressure data follows the power-law form previously observed
for (U,Th)Be$_{13}$, $C(T)=\gamma_s T+AT^3$ with $\gamma_s=0.64$ J/mol K
at $P=0.03$ kbar.  Previous low-temperature heat capacity measurements for
different Th concentrations \cite{Jin94} found $C(T)=\gamma_s T+AT^n$ with
a best-fit exponent $n$ near 4, an unphysical value, for Th concentrations
of 2.2\% and above. Our sample, with a slightly lower Th concentration,
retains the $T^3$ behavior. Indeed, this function proves to fit our
data well for all pressures. Three-parameter fits of $C(T)$ to the form
$\gamma_sT+AT^n$ give an exponent $n\approx 3$ at all pressures, as shown
in Figure \ref{f:gamma}.  With this in mind, we fix $n=3$ and carry
out two-parameter fits.  We find a steady increase in $\gamma_s$ with
pressure, while $A$ decreases more slowly; these quantities are shown in
Figure \ref{f:gamma}.

\begin{figure}
\begin{center}
\psfrag{DC (J/mol K2)}{\scalebox{2}{$\Delta C$ (J/mol K$^2$)}}
\scalebox{0.45}{\includegraphics{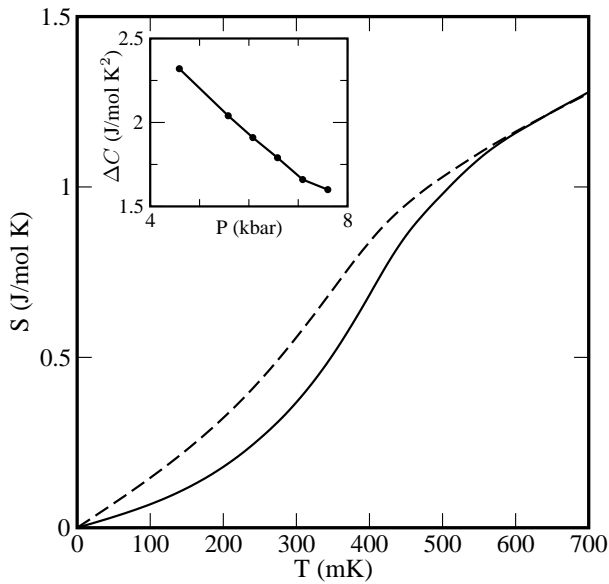}}
\caption{\small  
Entropy from 0 to 700 mK for 0.03 kbar (solid) and 5.58 kbar (dashed), the lowest
and highest pressures in Figure \ref{f:CvsT}.  Inset: Specific heat jump at the
transition, for single-transition regime.}
\label{f:entropy}
\end{center}
\end{figure}

As a further check, we extrapolate the specific heat curves to $T=0$
according to the above fits.  We then integrate to find the entropy $S$
by $S(T)=\int_0^T dT\frac{C(T)}{T}$.  Figure \ref{f:entropy} compares the
entropy for the lowest and highest pressure curves of Figure \ref{f:CvsT}.
By 700 mK, safely in the normal state, the total
entropy varies by only a few percent among pressures.  The extra entropy
under the $C/T$ peak at low pressures offsets the extra entropy under
the low-temperature tail at high pressures.

With this confirmation, we return to the $\gamma_s$. The magnitude of
the change with pressure is striking, more than  a factor of two.  Furthermore,
the change is completely reversible, even without raising the temperature above
300 mK.  In fact, exploring the reversibility of the low-temperature heat
capacity originally motivated our measurements. The phase diagram of
U$_{1-x}$Th$_x$Be$_{13}$ includes boundaries within the superconducting regime as
functions of concentration and pressure, as well as temperature.  For Th
concentrations with two transitions, $\mu$SR measurements finds weak local
magnetic order below the lower-temperature transition \cite{Heffner90,
Heffner89}.  No local order appears for Th concentrations with a single
transition, suggesting a phase boundary near $x=0.02$ between the single phase at
$x<0.02$ and the lower-temperature phase for $x>0.02$.  Since increasing pressure
acts much like decreasing Th concentration \cite{Lambert, Sigrist}, pressure
measurements can cross an analogous phase boundary and explore its thermodynamic
properties.  

The earlier heat capacity measurements under pressure \cite{Zieve94} drove
the sample around rather than across the phase boundary: pressure was
always changed
at room temperature, with the sample then cooled into
one phase or the other.  Our bellows pressure cell allows us to
change pressure while cold, thereby crossing the transition directly.
We also use a sample with Th concentration closer to the transition
and extend the measurements to lower temperatures by using a dilution
refrigerator rather than the pumped ${}^3$He cryostat of the earlier work.

We use several paths through the pressure-temperature space to search for
hysteresis.  In one case, we change pressure, keeping temperature below 300
mK.  We then measure $C(T)$ from low temperature to above $T_c$.  After the
sample has warmed above $T_c$, we cool and repeat the specific heat
measurements from our lowest temperatures.  The two specific heat curves agree
to better than 0.5\% for both increasing and decreasing pressure, up to 5.5
kbar. Changing pressure at fixed temperature is less reliable, both because it
is more difficult to return exactly to the original pressure and because the
temperature always changes slightly during the pressure change, but again there
is no evidence of hysteresis in the specific heat.

We also find no evidence of latent heat when we monitor the sample
temperature while increasing or decreasing pressure.  The temperature
rises during any pressure change.  The effect is more noticeable upon
increasing pressure, but the difference is completely explained by
the heat load to the cryostat from adding additional room-temperature
helium and by the work done in compressing the sample.  We never find
a temperature reduction on changing pressure.  We conclude that any
transition with pressure in this region is second-order.

Although pressure might introduce additional defects into a
sample, perhaps even defects that could substantially alter the heat
capacity, such an explanation for the change in $\gamma_s$ also demands,
implausibly,
that the defects anneal away at low temperature. The large reversible
effect on $\gamma_s$ suggests that in fact the $\gamma_s$
is tied intimately to the mechanism of superconductivity itself.

Note that at our highest pressures $\gamma_s$ exceeds the normal-state
$C/T$ just above $T_c$.  This confirms other evidence of non-Fermi liquid
behavior in UBe$_{13}$. An entropy deficit in normal UBe$_{13}$ relative
to the superconducting phase has long been known, suggesting
that $\gamma_n(T)$ increases substantially below $T_c$.  Suppressing $T_c$
with a magnetic field bears this out, with $C/T$ increasing steadily
toward lower temperatures.  The same effect, with an even stronger
increase in $\gamma_n(T)$, appears for Th-doped UBe$_{13}$. Heat capacity in
a 3 T field shows $\gamma_n(T)$ of 1400 mJ/mole K$^2$ at 0.42 K.  To match
the measured superconducting entropy it must rise to 2300 mJ/mole K$^2$
at $T = 0$, a faster than linear increase in $\gamma_n(T)$ itself \cite{Kim91}.

The behavior of $\gamma_s$ emphasizes that increasing pressure and decreasing
Th concentration have analogous but not identical effects. As shown previously,
the topology of the phase diagram appears to be similar for the two variables,
but the temperature dependence of the transitions is not.  On decreasing Th
concentration below 2\%, $T_c$ rises.  While pressure also merges the
transitions, $T_c$ decreases monotonically.  Similarly, $\gamma_s$
generally {\em decreases} with decreasing Th concentration; it is more
an order of magnitude smaller in pure UBe$_{13}$ than for 2\% Th doping. Yet
$\gamma_s$ {\em increases} with increasing pressure.  

This suggests that changes in $\gamma_s$ come from quantitative rather
than qualitative changes in the order parameter. For a BCS superconductor,
the specific heat jump at the transition, $\Delta C$, and the magnitude
of the energy gap, $\Delta(0)$, satisfy $\Delta C=1.43\gamma_n T_c$ and
$\Delta(0)=1.76kT_c$.  Although these simple proportionalities fail for
a non-$s$-wave order parameter or strong coupling, the discontinuity is
still related to $\Delta(0)$. We fit our specific heat data with a single
sharp transition, with entropy conserved between our data and the fit.
The inset of Figure \ref{f:entropy} shows the size of the jump, at
pressures high enough that a single transition provides a good fit to
the data.  The jump decreases with increasing pressure, although not as
rapidly as $\gamma_s$ increases.

Other heavy-fermion materials also show
substantial changes in $\gamma_s$ with pressure.
In UPd$_2$Al$_3$, with superconducting $T_c$ near 1.5K and
antiferromagnetic $T_N\approx 18$K, $\gamma_s$ increases over 50\% at 10.8 kbar
of hydrostatic pressure \cite{Caspary}.  In this case the heat capacity
just above $T_c$, deep in the antiferromagnetic phase, increases by a comparable
amount.  A possible explanation is that separate electron subsystems are
responsible for the magnetic and superconducting behaviors, with the pressure
dependence arising only from the electrons responsible for the magnetism.

Another example is CeRhIn$_5$ \cite{Fisher02} at the boundary
between antiferromagnet and superconductor.  In this case, the
superconducting $\gamma_s$ vanishes at high pressures, but rises
steadily as pressure decreases from 21 to 15 kbar.  At 15 kbar, at the
antiferromagnetic transition, $\gamma_s$ has reached its value within
the antiferromagnet. The authors interpret the large linear term as
a consequence of finite regions of ungapped Fermi surface \cite{Fisher02}
that come from an increase in anisotropic impurity scattering near the
antiferromagnetic transition \cite{Hass}.

Thus a large and changing $\gamma_s$ appears not only in our
superconducting system but also in a superconducting antiferromagnet and
at a superconductor/antiferromagnet transition.  Whether there is any
further connection among these systems, such as proximity to an unrealized
antiferromagnetic transition in (U,Th)Be$_{13}$, remains to be seen.

In summary, we observe a large increase with pressure in $\gamma_s$, the linear
coefficient of the superconducting specific heat. We also find that all
pressure-dependent behavior is reversible, indicating that all phase transitions
in the region are second order.  The change in $\gamma_s$ with no change in the
impurity density appears inconsistent with some proposed explanations for the
origin of the linear term, including resonant impurity scattering. Whether or
not a non-zero $\gamma_s$ is itself an intrinsic property of (U,Th)Be$_{13}$,
its strong variation within a single sample is likely intrinsic and may prove a
useful signature of the nature of superconductivity in the material.

This work was supported by NSF under DMR-9733898 (UCD).  Work at Los Alamos
was performed under the auspices of the U.S. Department of Energy.

\end{document}